\documentclass[10pt,letterpaper]{article}

\usepackage{opex3}
\usepackage{amsfonts, amsmath, amssymb}
\usepackage{cite}

\begin{document}

\title{Jacobi photonic lattices and their SUSY partners}

\author{A. Z\'u\~niga-Segundo$^{1}$, B. M. Rodr\'{\i}guez-Lara$^{\ast 2}$, David J. Fern\'andez C.$^{3}$, and H. M. Moya-Cessa$^{2}$ }

\address{$^{1}$Departamento de F\'{\i}sica, Escuela Superior de F\'{\i}sica y Matem\'{a}ticas, IPN Edificio 9 Unidad Profesional `Adolfo L\'{o}pez Mateos', 07738 M\'{e}xico D.F., M\'{e}xico.\\
$^{2}$Instituto Nacional de Astrof\'{i}sica, \'{O}ptica y Electr\'{o}nica \\ Calle Luis Enrique Erro No. 1, Sta. Ma. Tonantzintla, Pue. CP 72840, M\'{e}xico.\\
$^{3}$Departamento de F\'{\i}sica, CINVESTAV, A. P. 14-740, 07000 M\'{e}xico D.F., M\'{e}xico.}

\email{$^{\ast}$ bmlara@inaoep.mx}

\begin{abstract}
We present a classical analog of quantum optical deformed oscillators in arrays of waveguides. 
The normal modes of these one-dimensional photonic crystals are given in terms of Jacobi polynomials.
We show that it is possible to attack the problem via factorization by exploiting the corresponding quantum optical model. 
This allows us to provide an unbroken supersymmetric partner of the proposed Jacobi lattices.
Thanks to the underlying $SU(1,1)$ group symmetry of the lattices, we present the analytic propagators and impulse functions for these one-dimensional photonic crystals.
\end{abstract}

%\pacs{42.25.Bs, 11.30.Pb, 24.81.Qb, 42.82.Et }
\ocis{ (230.4555) Coupled Resonators;  (230.5298) Photonic Crystals; (230.7370) Waveguides;  (350.5500) Propagation.}

%%%%%%%%%%%%%%%%%%%%%%%%%%%%%%%%%%%%%%%%%%%%%%%%%%%%%%%%%%%%%%%%%%%%%%%%%%%%%%%%%%%%%%%%%%%%%%%%
%\bibliographystyle{osajnl}
%\bibliography{D:/ExternalHD/Bibliography/references}

%%%%%%%%%%%%%%%%%%%%%%%%%%%%%%%%%%%%%%%%%%%%%%%%%%%%%%%%%%%%%%%%%%%%%%%%%%%%%%%%%%%%%%%%%%%%%%%%

\section{Introduction} \label{sec:S1}

Supersymmetry (SUSY) was first used as a way to unify bosonic and fermionic sectors in string models and, along the time, it has been used to unify space-time and internal symmetries in high energy physics, to generalize gravity in relativistic physics, to find and categorize analytically solvable potentials in quantum mechanics, just to mention a few examples \cite{Cooper1995p267,Fernandez2005p236,Fernandez2010p3}. 
In optics, planar waveguides with some particular refractive index profiles have been shown to accept isospectral partners in the paraxial regime \cite{Chumakov1994p51} and SUSY has provided a method to generate a family of isospectral potentials to optimize quantum cascade lasers \cite{Tomic1997p1033,Bai2006p4043}.
In quantum optics, isospectral partners for ion-trap Hamiltonians have been used to propose entanglement generation by adiabatic ground-state transition \cite{Unanyan2003p133601},  the simulation of the Wess-Zumino SUSY model in 2+1 dimensions with cold atom-molecule mixtures in the presence of a two-dimensional optical lattice \cite{Yu2010p150605} and that of the electric dipole moment of neutral relativistic particles related to SUSY models in ion-traps setups \cite{Tenev2013p022103}.

Optical analogies of quantum systems realized in waveguide arrays have recently impacted the field of integrated optical structures \cite{ElGanainy2013p161105}. 
In particular, SUSY photonic lattices can be used to provide phase matching conditions between large number of modes allowing the pairing of isospectral crystals \cite{Miri2013p233902,Valle2013p40008}.
In the following, we provide a class of photonic lattices showing discrete-SUSY and find their analytic spectrum, propagator and impulse function. 
Our one-dimensional crystals are also the classical analog to a class of generalized deformed oscillators \cite{Daskaloyannis1991p789,Dodonov1998p469} and can simulate non-classical squeezed light with the propagation of classical light \cite{Sukhorukov2013p053823}.
The normal modes coefficients of these waveguide arrays are given in terms of Jacobi polynomials and this the origin of their denomination.
We follow a factorization method \cite{Infeld1951p21} to study our Jacobi lattices  and to propose a feasible SUSY parter for them.
The analogy between the quantum and the classical optics systems allows us to present a closed form for the propagator and impulse function of our photonic crystals.

%%%%%%%%%%%%%%%%%%%%%%%%%%%%%%%%%%%%%%%%%%%%%%%%%%%%%%%%%%%%%%%%%%%%%%%%%%%%%%%%%%%%%%%%%%%%%%%%
\section{Jacobi lattices} \label{sec:S2}
\begin{figure}
\centerline{\includegraphics[scale=1]{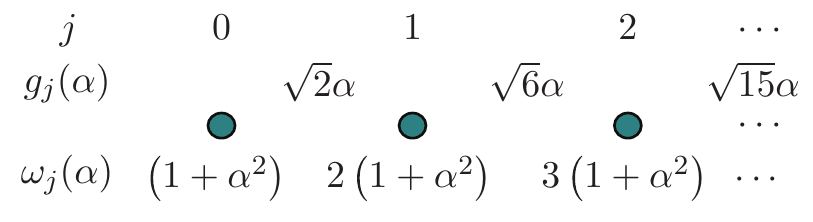}}
\caption{The effective refractive index, $\omega_{j}(\alpha)$, and coupling, $g_{j}(\alpha)$, functions for the Jacobi lattices.}  \label{fig:Fig1}
\end{figure}

\begin{figure}
\centerline{\includegraphics[scale=1]{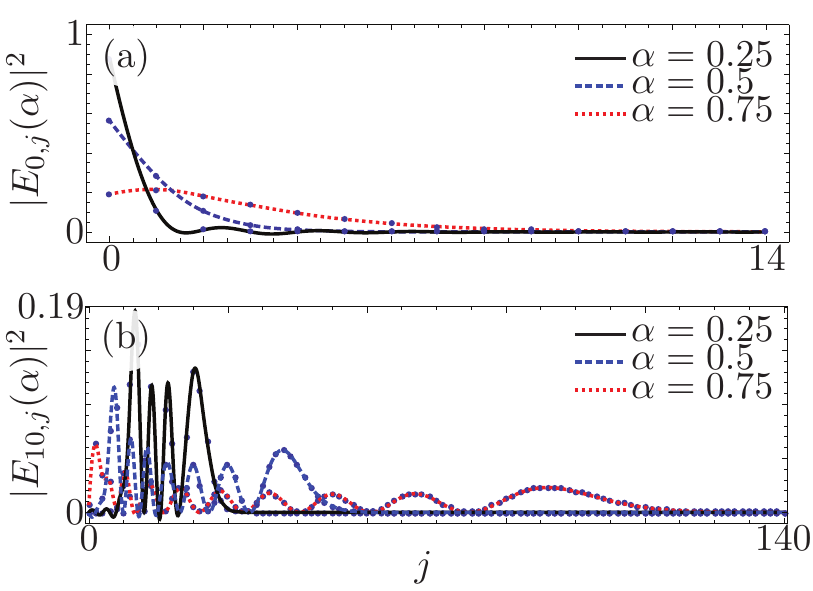}}
\caption{(Color online)The squared amplitudes corresponding to the (a) ground, $k=0$,  and  (b) tenth, $k=10$, normal mode, $\vert \alpha_{k} \rangle = \sum_{j=0}^{\infty} E_{k,j}(\alpha) \vert j \rangle$ for different values of the deformation parameter $\alpha$. The lines are a fit with a polynomial of order $20$.}  \label{fig:Fig2}
\end{figure}

%%%%%%%%%%%%%%%%%%%%%%%%%%%%%%%%%%%%%%%%%%%%%%%%%%%%%%%%%%%%%%%%%%%%%%%%%%%%%%%%%%%%%%%%%%%%%%%%
Let us consider the Sch\"odinger-like equation for a generalized deformed oscillator \cite{Daskaloyannis1991p789},
\begin{eqnarray}
\left[ i \partial_{z} + (1+\alpha^2)(\hat{n}+1) - \alpha\left(\sqrt{\hat{n}+1}~\hat{a}^\dagger + \hat{a} ~\sqrt{\hat{n}+1}\right) \right] \vert \psi \rangle = 0, \label{eq:SchEq}
\end{eqnarray}
where the shorthand notation $\partial_{z}$ stands for derivation with respect to $z$, the creation (ahhihilation) and number operators are given by $\hat{a}^{\dagger}$ ($\hat{a}$) and $\hat{n}$, respectively, and the parameter $\alpha$ is a real number that characterizes the deformed oscillator such that $\alpha \ne \pm 1$.
The action of the operators over a Fock state, also known as number state $\vert j \rangle$, is given by  $\hat{a}^{\dagger} \vert j \rangle = \sqrt{j+1} \vert j+1 \rangle$, $\hat{a} \vert j \rangle = \sqrt{j} \vert j -1 \rangle$ and  $\hat{n} \vert j \rangle = j \vert j \rangle$.
This allows us to write any given state of the system as the superposition $\vert \psi \rangle = \sum_{j=0}^{\infty} \mathcal{E}_{j} \vert j \rangle$ leading to the differential equation set 
\begin{eqnarray}
i\partial_{z} \mathcal{E}_{j} + (1+\alpha^2)(j+1) \mathcal{E}_j - \alpha \sqrt{(j+1)(j+2)} \mathcal{E}_{j+1} - \alpha \sqrt{j(j+1)} \mathcal{E}_{j-1}=0, \quad \mathcal{E}_{-1}=0. \label{eq:DiffSet}
\end{eqnarray}
This differential set can be related to an array of photonic waveguides as shown in Fig. \ref{fig:Fig1} \cite{Jones1965p261,Christodoulides2003p817}, such that we have a classical analog for our quantum optics system.
The experimental realization of this class of photonic lattices is feasible; cf. a discussion on a similar type of refractive index scaling and coupling given in \cite{RodriguezLara2011p053845,PerezLeija2012p013848,RodriguezLara2013p12888} and the fact that our lattices are a particular subclass of those given in \cite{Sukhorukov2013p053823} with parameters $N=1$, $\beta_{i}= 1+ \alpha^{2}$ and $\beta_{s}=0$.
In order to study our waveguide array, we will take advantage of the quantum optics model and follow an algebraic approach \cite{Daskaloyannis1991p789}.
For this reason, we can rewrite the Schr\"odinger-like equation in (\ref{eq:SchEq}) as $- i \partial_{z} \vert \psi \rangle = \hat{H} \vert \psi \rangle $  with the Hamiltonian,
\begin{eqnarray}
\hat{H} &=& (1+\alpha^2) \hat{K}_{0} - \alpha \left( \hat{K}_{+}  + \hat{K}_{-} \right), \label{eq:SU11Ham}
\end{eqnarray}
in terms of the elements of the $SU(1,1)$ group, $\hat{K}_{0} = \hat{n} +1$, $\hat{K}_{+} = \sqrt{\hat{n} +1} ~\hat{a}^{\dagger}$ and $\hat{K}_{-} = \hat{a} \sqrt{\hat{n} +1}$, that fulfill $\left[\hat{K}_{0}, \hat{K}_{\pm} \right]= \pm \hat{K}_{\pm}$ and $\left[\hat{K}_{+}, \hat{K}_{-} \right]= -2 \hat{K}_{0}$.
It is straightforward to diagonalize the expression in (\ref{eq:SU11Ham}) by use of the rotation $\hat{R}(\xi) = e^{-\frac{\xi}{2} \left( \hat{K}_{+}  - \hat{K}_{-}  \right)}$, 
\begin{eqnarray}
\hat{H}_{R} &=& \hat{R}(\xi) \hat{H} \hat{R}(-\xi), \\
&=& \left[(1+\alpha^2)\cosh\xi-2\alpha\sinh\xi\right] \hat{K}_{0}, \\
&=& \left( 1 - \alpha^{2} \right)\hat{K}_{0},
\end{eqnarray}
where we have used $\tanh \xi = 2  \alpha / ( 1 + \alpha^{2})$. 
Thus, the spectrum of our photonic lattice is given by
\begin{eqnarray}
\Omega_{k}(\alpha) = \left( 1 - \alpha^{2} \right)(k+1), \quad k= 0, 1, 2, \ldots \label{eq:Sp}
\end{eqnarray}
The corresponding normal modes, $\vert \alpha_{k} \rangle = \hat{R}(-\xi) \vert k \rangle$, are given by a rotation over the basis in the diagonal representation, given by the number states $\vert k \rangle$, and can be reduced to the form 
\begin{eqnarray}
\vert \alpha_{k} \rangle &=& \sum_{j=0}^{\infty} \sqrt{\frac{k+1}{j+1}}(-1)^k(1-\alpha^2){\alpha}^{k-j}P^{(1,k-j)}_j(2\alpha^2-1) \vert j \rangle, \label{eq:NM}
\end{eqnarray}
in terms of the Jacobi polynomials $P^{\gamma, \beta}_{n}(x)$ \cite{Abramowitz1970}.
It is for this reason that we christen our waveguide arrays as Jacobi lattices.
Figure \ref{fig:Fig2} shows the squared amplitudes of the ground and tenth normal mode for different values of the deformation parameter.

%%%%%%%%%%%%%%%%%%%%%%%%%%%%%%%%%%%%%%%%%%%%%%%%%%%%%%%%%%%%%%%%%%%%%%%%%%%%%%%%%%%%%%%%%%%%%%%%
%%%%%%%%%%%%%%%%%%%%%%%%%%%%%%%%%%%%%%%%%%%%%%%%%%%%%%%%%%%%%%%%%%%%%%%%%%%%%%%%%%%%%%%%%%%%%%%%

%%%%%%%%%%%%%%%%%%%%%%%%%%%%%%%%%%%%%%%%%%%%%%%%%%%%%%%%%%%%%%%%%%%%%%%%%%%%%%%%%%%%%%%%%%%%%%%%
%%%%%%%%%%%%%%%%%%%%%%%%%%%%%%%%%%%%%%%%%%%%%%%%%%%%%%%%%%%%%%%%%%%%%%%%%%%%%%%%%%%%%%%%%%%%%%%% 

As we have shown in the past for a photonic lattice with $SU(2)$ symmetry \cite{RodriguezLara2013}, we can follow an equivalent algebraic approach with the $SU(1,1)$ group and calculate the propagator function by using the quantum optics analogy:
\begin{eqnarray}
\vert \psi(z) \rangle = e^{f(z) \hat{K}_{+}} e^{- 2 \ln g(z) ~\hat{K}_{0}} e^{f(z) \hat{K}_{-}} \vert \psi(0) \rangle
\end{eqnarray}
with
\begin{eqnarray}
f(z)&=&\alpha\left( \frac{1 - e^{-i(1-\alpha^2)z}}{1 - \alpha^2 e^{-i(1-\alpha^2)z}} \right), \\
g(z)&=&\cos\left[{{\frac{1}{2}}(1-\alpha^2)z}\right]-i\left({\frac{\alpha^2+1}{\alpha^2-1}}\right) \sin\left[{{\frac{1}{2}}(1-\alpha^2)z}\right], 
\end{eqnarray}
where $\vert \psi(z) \rangle = \sum_{j=0}^{\infty} \mathcal{E}_{j}(z) \vert j \rangle$ is a vector containing the information of the propagating classical field amplitude at the $j$th waveguide, $\mathcal{E}_{j}(z)$.
Thus, the classical field amplitude at the $j$th waveguide for an initial field impinging just the $k$th waveguide, also known as impulse function, 
\begin{eqnarray}
I_{j,k}(z) &=& \sqrt{\frac{k+1}{j+1}}  ~\frac{f^{k-j}(z)}{g^{2(j+1)}(z)} \left[ \frac{2}{h(z)-1} \right]^{j}  P^{(1,k-j)}_{j}\left[ h(z) \right], \label{eq:ImpulseSU11}
\end{eqnarray}
is given in terms of Jacobi polynomials and the auxiliary function
\begin{eqnarray} 
h(z)&=& 1-\frac{2 \left(\alpha ^2-1\right)^2}{\alpha ^4-2 \alpha ^2 \cos \left[ \left(1-\alpha ^2\right) z\right] + 1}.
\end{eqnarray}

\begin{figure}
\centerline{\includegraphics[scale=1]{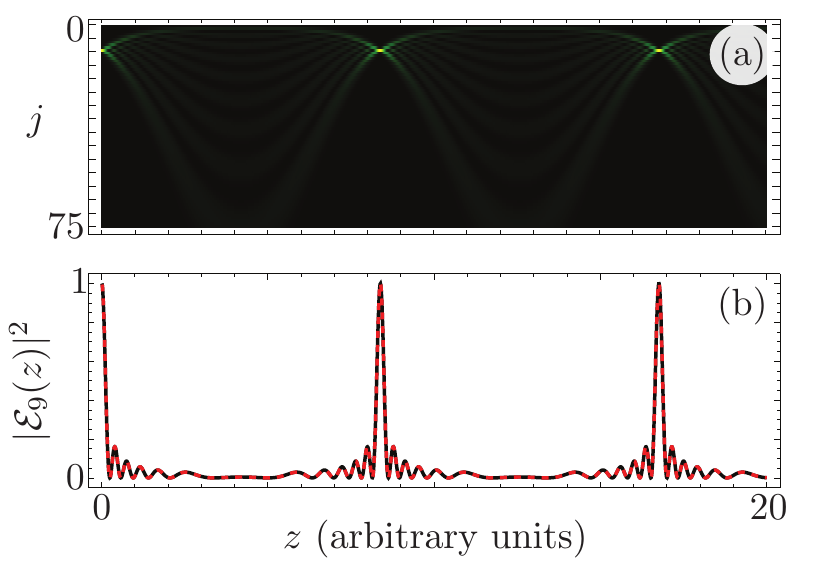}}
\caption{(Color online) (a) Numerical propagation of light intensity for an initial field impinging at the $j=9$ waveguide of a Jacobi lattice with a deformation parameter $\alpha=0.5$. (b) Comparison of the numerical (solid black) and theoretical (dotted red) intensities at the $j=9$ waveguide.}  \label{fig:Fig3}
\end{figure}

Note that as (\ref{eq:SU11Ham}) is a compact operator we will expect coherent oscillations for a single waveguide input as shown in Fig. \ref{fig:Fig3}(a) for an initial classical field impinging at the tenth waveguide, $j=9$, of a Jacobi array with parameter $\alpha = 0.5$; although (\ref{eq:SU11Ham}) is composed by compact, $\tilde{K}_{0}$, and non-compact, $(\tilde{K}_{+} + \tilde{K}_{-})$, members of $SU(1,1)$ the relation $ 1 + \alpha^{2} \ge \vert \alpha \vert$ makes it a compact operator \cite{Puri1996p1563}, 
The numerical and theoretical results are in good agreement, Fig. \ref{fig:Fig3}(b), with differences of the order of $10^{-15}$.
We used a lattice size of 200 waveguides in the numerical simulations, which is in the experimental range, and the light intensity at the last waveguide was of the order of $10^{-17}$ for the example shown in Fig. \ref{fig:Fig3}.

%%%%%%%%%%%%%%%%%%%%%%%%%%%%%%%%%%%%%%%%%%%%%%%%%%%%%%%%%%%%%%%%%%%%%%%%%%%%%%%%%%%%%%%%%%%%%%%%
%%%%%%%%%%%%%%%%%%%%%%%%%%%%%%%%%%%%%%%%%%%%%%%%%%%%%%%%%%%%%%%%%%%%%%%%%%%%%%%%%%%%%%%%%%%%%%%% 
 
%%%%%%%%%%%%%%%%%%%%%%%%%%%%%%%%%%%%%%%%%%%%%%%%%%%%%%%%%%%%%%%%%%%%%%%%%%%%%%%%%%%%%%%%%%%%%%%%
\section{SUSY partner for Jacobi lattices} \label{sec:S3}

\begin{figure}
\centerline{\includegraphics[scale=1]{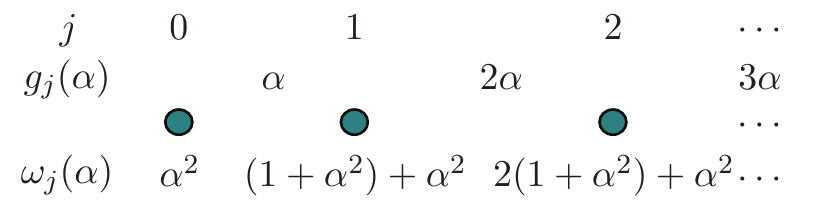}}
\caption{The effective refractive index, $\omega_{j}(\alpha)$, and coupling, $g_{j}(\alpha)$, functions for the SUSY partner of our Jacobi lattices.}  \label{fig:Fig4}
\end{figure}

\begin{figure}
\centerline{\includegraphics[scale=1]{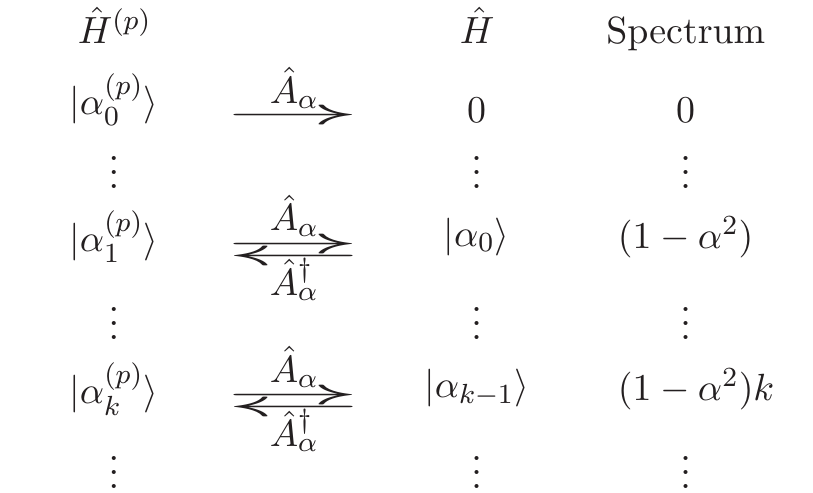}}
\caption {Relation between the normal modes of the Jacobi lattices, $\vert \alpha_{k} \rangle$ with spectrum $(1 - \alpha^{2})(k+1)$, and the normal modes of their susy partner, $\vert \alpha_{k}^{(p)} \rangle$ with spectrum $(1 - \alpha^{2})k$.}  \label{fig:Fig5}
\end{figure}

By defining deformed, nonlinear, parametrized creation and anihilation operators, 
\begin{eqnarray}
\hat{A}^\dagger_\alpha = \hat{a}^{\dagger} -\alpha\sqrt{\hat{n}+1}, \quad \hat{A}_{\alpha} = \hat{a}-\alpha\sqrt{\hat{n}+1},  
\end{eqnarray}
it is possible to factorize (\ref{eq:SchEq}) into
\begin{eqnarray}
\left[ i \partial_{z} + \hat{A}_{\alpha}  \hat{A}^{\dagger}_{\alpha} \right] \vert \psi \rangle = 0.
\end{eqnarray}
Now, it is straightforward to propose a SUSY partner \cite{Cooper1995p267} to our Jacobi lattice in the form:
\begin{eqnarray}
\left[ i \partial_{z} +  \hat{A}^{\dagger}_{\alpha} \hat{A}_{\alpha}  \right] \vert \psi \rangle = 0.
\end{eqnarray}
This leads to an array of waveguides described by the differential set
\begin{eqnarray}
i\partial_{z} \mathcal{E}_{j} + \left[ (1+\alpha^2)j + \alpha^2 \right] \mathcal{E}_j - \alpha (j+1) ~\mathcal{E}_{j+1} - \alpha j ~\mathcal{E}_{j-1}=0. \quad \mathcal{E}_{-1}=0,
\end{eqnarray}
and shown in Fig. \ref{fig:Fig4}.
Here, we have to use an alternative representation for the elements of the $SU(1,1)$ group: $\tilde{K}_{0} = \hat{n} + 1/2$, $\tilde{K}_{+} = \sqrt{\hat{n}} ~\hat{a}^{\dagger}$ and $\tilde{K}_{-} = \hat{a} \sqrt{\hat{n}}$  such that the equivalent partner Hamiltonian for the quantum optics analog is given by
\begin{eqnarray}
\hat{H}_{p}&=& (1+\alpha^2) \tilde{K_{0}} - \alpha \left( \tilde{K}_{+}  + \tilde{K}_{-}  \right)  - \frac{1}{2} \left(1 -  \alpha^2 \right) . \label{eq:SU11HamPartner}
\end{eqnarray}
Otherwise, the procedure to find the spectrum and normal modes is identical to that in the past section; we define a rotation $\tilde{R} = e^{- \xi \left( \tilde{K}_{+} - \tilde{K}_{-} \right)/2 }$ such that diagonalizes the quantum analog Hamiltonian,
\begin{eqnarray}
\hat{H}_{pR} &=& \tilde{R}(\xi) \hat{H}_{p} \tilde{R}(-\xi), \\
&=& \left[(1+\alpha^2)\cosh\xi-2\alpha\sinh\xi\right] \tilde{K}_{0}  - \frac{1}{2} \left(1 -  \alpha^2 \right) , \\
&=& \left( 1 - \alpha^{2} \right) \left( \tilde{K}_{0} - \frac{1}{2} \right) , 
\end{eqnarray}
where we have used $\tanh \xi = 2  \alpha / ( 1 + \alpha^{2})$ again. 
In this case the spectrum is identical, up to an extra first term, to that in (\ref{eq:Sp}),
\begin{eqnarray}
\Omega_{k}^{(p)}(\alpha) = \left( 1 - \alpha^{2} \right) k  , \quad k=0,1,2, \ldots \label{eq:SpPartner}.
\end{eqnarray}
The corresponding eigenmodes can be written as:
\begin{eqnarray}
\vert\alpha^{(p)}_{k} \rangle = \sum_{j=0}^{\infty}  \sqrt{1-\alpha^2}(-1)^k {\alpha}^{k-j} P^{(0,k-j)}_j(2\alpha^2-1) \vert j \rangle \label{eq:NMPartner}
\end{eqnarray}
where Jacobi polynomials of a lower order than before appear; these can be reduced to the ordinary hypergeometric function $P^{(0,k-j)}_j(2\alpha^2-1)=~_{2}F_{1}(-j,k+1,1,1-\alpha^2)$ \cite{Abramowitz1970}.
Note that the lower energy normal mode is annihilated as expected from the annihilation operator, $\hat{A}_{\alpha} \vert \alpha_{k}^{(p)} \rangle =0$, and we can classify the SUSY as unbroken \cite{Cooper1995p267}.
Using the properties of Jacobi polynomials \cite{Gradshteyn2007}, it possible to show that the rest of the normal modes of the isospectral partner are related to the normal modes of the Jacobi lattice as
\begin{eqnarray}
\hat{A}_{\alpha} \vert \alpha_{k}^{(p)} \rangle &=& \sqrt{k \left( 1 - \alpha^{2} \right)} ~\vert \alpha_{k-1} \rangle , \quad k\ge 1, \\
\hat{A}_{\alpha}^{\dagger} \vert \alpha_{k} \rangle &=& \sqrt{(k+1) \left( 1 - \alpha^{2} \right)} ~ \vert \alpha_{k+1}^{(p)} \rangle , \quad k\ge 0.
\end{eqnarray}
Figure \ref{fig:Fig5} shows a diagram approach relating the normal modes of both photonic lattices and their spectra.

%%%%%%%%%%%%%%%%%%%%%%%%%%%%%%%%%%%%%%%%%%%%%%%%%%%%%%%%%%%%%%%%%%%%%%%%%%%%%%%%%%%%%%%%%%%%%%%%
%%%%%%%%%%%%%%%%%%%%%%%%%%%%%%%%%%%%%%%%%%%%%%%%%%%%%%%%%%%%%%%%%%%%%%%%%%%%%%%%%%%%%%%%%%%%%%%%

%%%%%%%%%%%%%%%%%%%%%%%%%%%%%%%%%%%%%%%%%%%%%%%%%%%%%%%%%%%%%%%%%%%%%%%%%%%%%%%%%%%%%%%%%%%%%%%%
%%%%%%%%%%%%%%%%%%%%%%%%%%%%%%%%%%%%%%%%%%%%%%%%%%%%%%%%%%%%%%%%%%%%%%%%%%%%%%%%%%%%%%%%%%%%%%%%

\begin{figure}
\centerline{\includegraphics[scale=1]{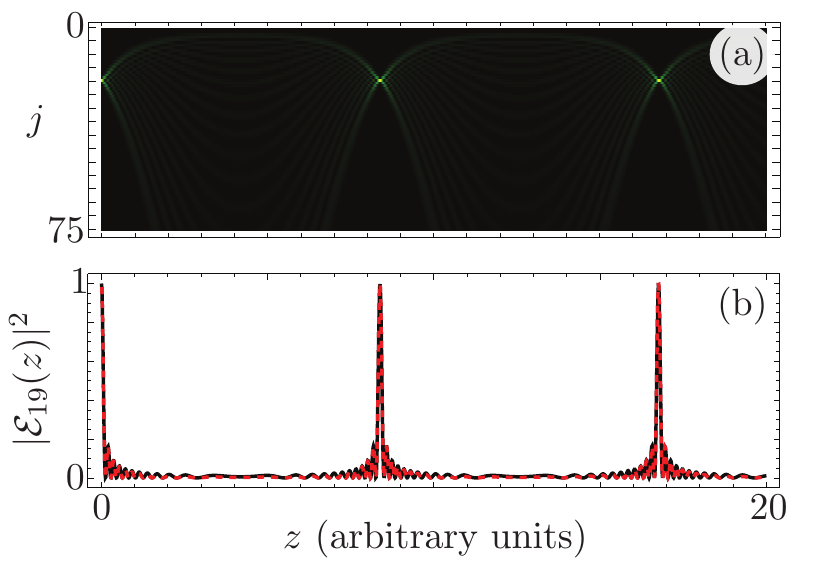}}
\caption {(Color online) (a) Numerical propagation of light intensity for an initial field impinging at the $j=9$ waveguide of a Jacobi SUSY partner lattice with a deformation parameter $\alpha=0.5$. (b) Comparison of the numerical (solid black) and theoretical (dotted red) intensities at the $j=19$ waveguide.}  \label{fig:Fig6}
\end{figure}

The propagator for this isospectral partner of our Jacobi lattices is given by
\begin{eqnarray}
\vert \psi(z) \rangle = e^{-\frac{i z}{2}(1-\alpha^2)} e^{f(z) \tilde{K}_{+}} e^{- 2 \ln g(z) ~\tilde{K}_{0}} e^{f(z) \tilde{K}_{-}} \vert \psi(0) \rangle,
\end{eqnarray}
where the auxiliary functions $f(z)$ and $g(z)$ are identical to those for the original Jacobi lattices and the corresponding impulse function is
\begin{eqnarray}
I_{j,k}(z) &=&  e^{-\frac{i z}{2}(1-\alpha^2)}  ~\frac{f^{k-j}(z)}{g^{2(j+1)}(z)} \left[ \frac{2}{h(z)-1} \right]^{j}  P^{(0,k-j)}_{j}\left[ h(z) \right]. \label{eq:ImpulsePartner}
\end{eqnarray}
Figure \ref{fig:Fig6}(a) shows the intensity from a numerical propagation of a beam of light impinging the twentieth, $j=19$, waveguide of a Jacobi SUSY partner lattice with deformation parameter $\alpha = 0.5$. 
In this simulation the lattice was composed by 200 waveguides and the field intensity at the last waveguide was always smaller than $7 \times 10^{-6}$.
We compare the squared impulse function versus the numerical intensity at the $j=19$ waveguide in Fig. \ref{fig:Fig6}(b) where we can see the expected coherent oscillation of the field amplitude. 

%%%%%%%%%%%%%%%%%%%%%%%%%%%%%%%%%%%%%%%%%%%%%%%%%%%%%%%%%%%%%%%%%%%%%%%%%%%%%%%%%%%%%%%%%%%%%%%%
%
%%%%%%%%%%%%%%%%%%%%%%%%%%%%%%%%%%%%%%%%%%%%%%%%%%%%%%%%%%%%%%%%%%%%%%%%%%%%%%%%%%%%%%%%%%%%%%%%
%%%%%%%%%%%%%%%%%%%%%%%%%%%%%%%%%%%%%%%%%%%%%%%%%%%%%%%%%%%%%%%%%%%%%%%%%%%%%%%%%%%%%%%%%%%%%%%%

%%%%%%%%%%%%%%%%%%%%%%%%%%%%%%%%%%%%%%%%%%%%%%%%%%%%%%%%%%%%%%%%%%%%%%%%%%%%%%%%%%%%%%%%%%%%%%%%
\section{Conclusions} \label{sec:S4}
%%%%%%%%%%%%%%%%%%%%%%%%%%%%%%%%%%%%%%%%%%%%%%%%%%%%%%%%%%%%%%%%%%%%%%%%%%%%%%%%%%%%%%%%%%%%%%%%

We have taken advantage of factorization techniques in quantum mechanics to propose two classical isospectral photonic lattices such that their quantum optics analogies are defined by the Hamiltonians $\hat{H}= \hat{A}_{\alpha} \hat{A}_{\alpha}^{\dagger}$ and its SUSY partner $\hat{H}^{(p)}= \hat{A}_{\alpha}^{\dagger} \hat{A}_{\alpha}$ with $\alpha \in \mathbb{R}$ and $\alpha \ne \pm 1$.
These Hamiltonians are related in such a way, $ \hat{H} \hat{A}_{\alpha} = \hat{A}_{\alpha} \hat{H}_{p}$ and  $  \hat{A}_{\alpha}^{\dagger} \hat{H}= \hat{H}_{p}\hat{A}_{\alpha}^{\dagger}$, that their spectra are identical, $\Omega_{k}(\alpha) = \Omega_{k+1}^{(p)}(\alpha)$, and their normal modes are related through the deformed creation (annihilation) operators, $\vert \alpha_{k-1} \rangle = \left[k \left( 1 - \alpha^{2} \right)\right]^{-1/2}  \hat{A}_{\alpha} \vert \alpha_{k}^{(p)} \rangle$ for $k\ge 1$ and $\vert \alpha_{k}^{(p)} \rangle = \left[(k+1) \left( 1 - \alpha^{2} \right)\right]^{-1/2}  \hat{A}_{\alpha}^{\dagger} \vert \alpha_{k} \rangle$.
The two waveguide arrays correspond to an unbroken SUSY because the annihilation operator annihilates the ground state of the partner, $\hat{A}_{\alpha} \vert \alpha_{0}^{(p)} \rangle = 0$.
We have named these arrays of waveguides Jacobi lattices because the amplitudes of their normal modes are given in terms of Jacobi polynomials. 
The arrays are feasible of experimental realization as the individual waveguides require a linearly increasing refractive index while the couplings between waveguides follow a square root distribution in the Jacobi lattice and increase linearly in the SUSY partner.
These lattices show a $SU(1,1)$ symmetry with group parameter $k=1$ for the Jacobi lattice and $k=1/2$ for its partner that allowed us to present an analytic propagator for them.
The analog Hamiltonian operators of our photonic lattices are compact, this guarantees that the impulse functions will show coherent oscillations and we could keep the size of the lattices finite depending on the initial field amplitudes.

%%%%%%%%%%%%%%%%%%%%%%%%%%%%%%%%%%%%%%%%%%%%%%%%%%%%%%%%%%%%%%%%%%%%%%%%%%%%%%%%%%%%%%%%%%%%%%%%
\section*{Acknowledgments}
HMMC is grateful to P. Aleahmad for fruitful discussions.

%%%%%%%%%%%%%%%%%%%%%%%%%%%%%%%%%%%%%%%%%%%%%%%%%%%%%%%%%%%%%%%%%%%%%%%%%%%%%%%%%%%%%%%%%%%%%%%%

\end{document}